\documentclass[runningheads]{llncs}

\usepackage{epsfig}
\usepackage{subcaption}
\usepackage{calc}
\usepackage{amssymb}
\usepackage{amstext}
\usepackage{amsmath}

\usepackage[acronym]{glossaries}
\newacronym{eeg}{EEG}{electroencephalographic data}
\newacronym{mi}{MI}{motor imagery}
\newacronym{bci}{BCI}{brain–computer interface}
\newacronym{snr}{SNR}{signal-to-noise ratio}

\newacronym{mrcp}{MRCP}{motor related cortical potential}

\newacronym{fbcsp}{FBCSP}{Filter-bank common spatial pattern}
\newacronym{csp}{CSP}{common spatial pattern}
\newacronym{vmd}{VMD}{Visual Mode Decomposition}
\newacronym{stft}{STFT}{Short Time Fourier Transform}
\newacronym{psd}{PSD}{Power Spectral Density}
\newacronym{dtw}{DTW}{Dynamic Time Warping}

\newacronym{ml}{ML}{machine learning}
\newacronym{dl}{DL}{deep learning}
\newacronym{cnn}{CNN}{convolutional neural networks}
\newacronym{ae}{AE}{autoencoder}
\newacronym{vae}{VAE}{variational autoencoder}
\newacronym{ffnn}{FFNN}{feed-forward neural network}
\newacronym{elbo}{ELBO}{Evidence Lower Bound}
\newacronym{gan}{GAN}{generative adversarial network}
\newacronym{lda}{LDA}{linear discriminant analysis}
\newacronym{svm}{SVM}{support vector machine}
\newacronym{mse}{MSE}{mean square error}

\newacronym{dffn}{DFFN}{Densely Feature Fusion CNN}
\newacronym{cwcnn}{CW-CNN}{Channel-wise CNN}
\newacronym{tsgl}{TSGL-EEGNet}{Temporary Constrained Sparse Group Lasso enhanced EEGNet}
\newacronym{mbshallow}{MBShallow ConvNet}{Multibranch Shallow CNN}
\newacronym{qeegnet}{Q-EEGNet}{Quantized EEGNet}

\usepackage{blindtext}
\usepackage{makecell} 
\usepackage{siunitx} 
\usepackage{rotating}
\usepackage{balance}
\usepackage{lipsum}
\usepackage{hhline}
\usepackage{arydshln}
\usepackage{comment}
\usepackage{bm}
\usepackage{soul}
\usepackage{tabularray}

\DeclareMathOperator{\EX}{\mathbb{E}}

\usepackage[hyphens]{url}
\usepackage[hidelinks]{hyperref}
\hypersetup{breaklinks=true}
\usepackage[absolute]{textpos}
\usepackage{everyshi}

\begin{document}
\title{vEEGNet: learning latent representations to reconstruct EEG raw data via variational autoencoders}
%
%
\author{
    Alberto Zancanaro\inst{1}\orcidID{0000-0002-5276-7030} \and
    Giulia Cisotto\inst{1,2}\orcidID{0000-0002-9554-9367} \and
    Italo Zoppis\inst{2}\orcidID{0000-0001-7312-7123} \and
    Sara Lucia Manzoni\inst{2}\orcidID{0000-0002-6406-536X}
}
%
%
\institute{
    Department of Information Engineering, University of Padova,\\via Gradenigo 6/b, Padova, Italy \\
    \email{alberto.zancanaro.1@phd.unipd.it}\\
    \and
    Department of Informatics, Systems, and Communication, University of Milan-Bicocca,\\viale Sarca 336, Milan, Italy \\
    \email{\{giulia.cisotto, italo.zoppis, sara.manzoni\}@unimb.it}
    }

\maketitle              

\begin{abstract}
%
%
Electroencephalografic (EEG) data are complex multi~-~dimensional time~-~series which are very useful in many different applications, i.e., from diagnostics of epilepsy to driving brain-computer interface systems.
Their classification is still a challenging task, due to the inherent within- and between-subject variability as well as their low signal-to-noise ratio. On the other hand, the reconstruction of raw EEG data is even more difficult because of the high temporal resolution of these signals.
Recent literature has proposed numerous machine and deep learning models that could classify, e.g., different types of movements, with an accuracy in the range $70\%$ to $80\%$ (with 4 classes). On the other hand, a limited number of works targetted the reconstruction problem, with very limited results.
In this work, we propose vEEGNet, a DL architecture with two modules, i.e., an unsupervised module based on variational autoencoders to extract a latent representation of the multi-channel EEG data, and a supervised module based on a feed-forward neural network to classify different movements. Furthermore, to build the encoder and the decoder of VAE we exploited the well-known EEGNet network, specifically designed since 2016 to process EEG data.
We implemented two slightly different architectures of vEEGNet, thus showing state of the art classification performance, and the ability to reconstruct both low frequency and middle-range components of the raw EEG.
Although preliminary, this work is promising as we found out that the low-frequency reconstructed signals are consistent with the so-called motor-related cortical potentials, very specific and well-known motor-related EEG patterns and we could improve over previous literature by reconstructing faster EEG components, too.
Further investigations are needed to explore the potentialities of vEEGNet in reconstructing the full EEG data, to generate new samples, and to study the relationship between classification and reconstruction performance.
%
%
%

\end{abstract}



\begin{textblock*}{17cm}(1.7cm, 0.5cm)
\noindent\scriptsize This work has been submitted to the Springer Book Series "Communications in Computer and Information Science" for possible publication. Copyright may be transferred without notice, after which this version may no longer be accessible.\\
\end{textblock*}

\section{Introduction}

\textcolor{black}{
Classification and reconstruction of raw \gls{eeg} are critical tasks that \gls{ml}, and especially \gls{dl}, are asked to solve as this could help in many applications, e.g., from neuroscience to \gls{bci}, to support the diagnosis of neurological pathologies and the development of new rehabilitation protocols~\cite{EURASIP2022,WCOM2021,Frontiers2014}.
}
\textcolor{black}{
Both tasks involve solving a number of challenges, mostly relating to the high temporal resolution of the \gls{eeg} data, their inherent variability within- and between individuals, as well as their very low \gls{snr} \cite{ICC2013,ICC2020}.
The ability of a \gls{ml}/\gls{dl} models of extracting the most informative and compact representation of the rich and high-dimensional \gls{eeg} input dataset, i.e., a latent representation, plays a key role to both obtaining an effective classification and accurately reconstruct the data. Not only these two tasks, but also generation, compression, and denoising might be enabled by such models. 
}

\textcolor{black}{
The \gls{vae} is a variant of the standard auto-encoder model which consists of an effective encoding-decoding \gls{dl} architecture that projects data on a \emph{structured} latent space \cite{VAE_ORIGINAL_PAPER_kingma}. \gls{vae}s have been recently and successfully implemented for a number of unsupervised and semi-supervised learning problems, e.g., including latent feature extraction for further classification~\cite{VAE_ORIGINAL_PAPER_kingma,hinton2006reducing,li2019disentangled,MetroAgriFor2022}.
On the other hand, since 2016, \gls{cnn} have been successfully exploited for \gls{eeg} classification in the popular EEGNet architecture \cite{EEGNet_paper} and its several later variants. EEGNet is a 2-block architecture that implements both temporal and spatial convolutions and has been consistently shown to outperform earlier \gls{ml} models for \gls{eeg} classification \cite{FBCSP,zancanaro_CIBCB_article}.
}

\textcolor{black}{
Therefore, we directed our efforts in the investigation of the potentialities of the combination of \gls{vae} and EEGNet to reach good classification and reconstruction quality, at the same time, for \gls{eeg} data.
In our previous conference paper \cite{Zancanaro_2023_vEEGNet}, we proposed \emph{vEEGNet}, an architecture that combined a \gls{vae}, i.e., to extract artificial features, with a \gls{ffnn}, i.e., to classify the data into 4 different \gls{mi} classes (the imagination of a right or left hand, the movement of both feet or the tongue, as acquired in the public \emph{dataset 2a} from the BCI competition IV). We showed that vEEGNet can classify with accuracy values at the state of the art and, at the same time, can partially reconstruct the input \gls{eeg}, as limited to its low frequency components.
In the present work, we expand our contribution with two different implementations of vEEGNet, namely \emph{vEEGNet1} and \emph{vEEGNet2}, showing that vEEGNet2 has further reconstruction capabilities, i.e., it is able to reconstruct the middle-range components of the raw \gls{eeg}.
}

\textcolor{black}{The rest of this paper is organized as follows: Section~\ref{sec:state_of_the_art} reports the most relevant related work, Section~\ref{sec:methods} describes the \gls{vae} theory and introduces the general vEEGNet model. Section~\ref{sec:results} presents the two different implementations of vEEGNet, the dataset, the classification and reconstruction results with discussion w.r.t. the related state of the art. Finally, section~\ref{sec:conclusions} concludes the paper and paves the way toward new promising future directions.}

%
%
%
%
%

\section{State of the art}
\label{sec:state_of_the_art}

As \gls{eeg} is a very noisy data, with a large inherent variability across different subjects and within the same individual, developing a model which is able to reconstruct raw \gls{eeg} data with very high fidelity is a challenging task, as already shown by literature.
For this reason, most related work focused on the classification ability of \gls{ml} and \gls{dl} models, but rarely proposed models to specifically reconstructing such type of data.
A number of works suggested models and architectures with the primary objective of classifying e.g., movements, emotions, or epilectic seizures, with an inherent, however not deeply exploited, capability of reconstructing \gls{eeg}.
Thus, in this section we summarize the most relevant state of the art with respect to the aim of our work, i.e., both classification and reconstruction of \gls{eeg} signals during the imagination of movements (\gls{mi}) \cite{BIOSIGNALS2020}.

    %
\textcolor{black}{
%
There exist several models to classify \gls{eeg} of imagined movements. \gls{fbcsp}~\cite{FBCSP} is still the most common \gls{ml} model for classification, especially used in \gls{bci} applications. It consists of a 4-step algorithm involving a filter bank with multiple band-pass filters, a spatial filtering via \gls{csp}, a selection of the most relevant \gls{csp} features, and the classification using a standard \gls{ml} classifier (e.g., \gls{svm}). The well-established literature on this topic reports that this method can achieve up to $80\%$ accuracy in a 4-class \gls{mi} classification problem \cite{zancanaro_CIBCB_article,Sakhavi_AC_KS_TAB_4_5}.
%
Despite its popularity, \gls{fbcsp} has no reconstruction abilities. Therefore, its main role is barely as baseline comparison with more innovative \gls{dl} models.
Among others, \gls{cnn} have recently gained a lot of attention as effective architectures for classifying \gls{eeg} due to their ability to capture both temporal and spatial features. \gls{cnn} have been particularly used as building blocks for the very successful EEGNet architecture. EEGNet has been presented in 2016 by Lawhern et al.~\cite{EEGNet_paper} as an architecture to specifically classify raw \gls{eeg} signals (represented as multi-channel time-series). In its original version, it was made of 2 blocks, each one composed of 2 convolutional layers and a fully-connected layer, and it could barely reach an accuracy of about $70\%$. However, a number of variants and improvements of EEGNet have been later proposed, further outperforming both the baseline EEGNet as well as \gls{fbcsp}: among others, \gls{tsgl}~\cite{TSGL_EEGNet_AC_TAB_9}, \gls{mbshallow}~\cite{MBShallow_CovNet_AC_TAB_10}, \gls{mi}-EEGNet~\cite{MI_EEGNet_AC_TAB_10}, \gls{qeegnet}~\cite{qEEGNet}, \emph{DynamicNet}~\cite{zancanaro_CIBCB_article}, and other general-purpose \gls{cnn} models, namely \gls{cwcnn}~\cite{Sakhavi_AC_KS_TAB_4_5}, \gls{dffn}~\cite{DFNN_AC_TAB_7}, and the Monolithic Network~\cite{Monolitich_network_AC_KS_TAB_3}.
They implemented different pre-processing steps and some of them expanded the baseline EEGNet architecture with stacking multiple EEGNet units (e.g., TSGL-EEGNet and MBShallow ConvNet), finally reaching accuracy values slightly above $80\%$ in a 4-class \gls{mi} task.
}

%

To further extend the capabilities of \gls{dl} models to process \gls{eeg}, some recent architectures based on \gls{cnn} and \gls{ae} were proposed. They were shown to have both classification and basic reconstruction ability.
In~\cite{liu_2020_eeg_emotion_SAE_CNN}, the authors trained an architecture formed by a \gls{cnn} that extracted artificial features which were then fed as input to an \gls{ae}, with the aim of compressing and reconstructing \gls{eeg} signals corresponding to different emotions (the DEAP~\cite{koelstra_2012_DEAP_dataset} and the SEED~\cite{zheng_2015_SEED_dataset} public datasets were used). After the training, the optimized features were used as input to a \gls{ffnn}. The latter was used to classify low/high valence levels and low/high arousal levels, separately, in the DEAP dataset, achieving an accuracy of $89.49\%$ for the valence and of $92.86\%$ for the arousal, respectively (w.r.t. a chance level of about $50\%$). On the other hand, a 3-class classification problem (i.e., to discriminate positive, neutral, or negative emotional conditions) was the task for the \gls{ffnn} when applied to the SEED dataset, where it achieved an accuracy of $96.77\%$ (w.r.t. a chance level of about $33\%$).
In~\cite{yang_2018_DSAE_EEG}, the authors presented a denoising sparse autoencoder, i.e. an autoencoder that imposes a sparsity condition on the latent space, to learn features from \gls{eeg} for seizure detection. The learned features were then classified using a linear regression with very satisfactory classification results, i.e., $100\%$ accuracy on the dataset they used~\cite{yang_2018_DSAE_EEG_dataset}, both in binary and 3-class classification.
Despite the high classification performance and the use of an architecture with the potentiality of reconstructing the input data, the authors did not report any investigation on this aspect.
More recently, \cite{Bethge_2022_EEG2VEC} proposed EEG2VEC, a \gls{vae} based architecture that encoded emotions-related \gls{eeg} data. In this work, for the first time, the latent representation produced by a \gls{vae} was used to reconstruct \gls{eeg}. Interestingly, the authors succeeded in partially reconstructing the low-frequency shape of the signal, but not its entire frequency content, as well. In addition, the amplitude of the reconstructed data was about half that of the input one.
%
%
Incidentally, we also report that several works have recently used \gls{gan} to generate new synthetic \gls{eeg} samples of \gls{mi}, i.e., to increase the size of the \gls{eeg} datasets and, thus, the classification performance of \gls{dl}-based classifiers~\cite{luo2018eeggan,luo_2020_reconstruction_with_gan,tian_2023_Dual_encoder_VAE_GAN_generation}.
However, even though interesting and promising, this line of literature mainly addressed the classification problem, providing a substantially different approach from the one that is proposed in this work.

Thus, since our previous work~\cite{Zancanaro_2023_vEEGNet}, we have directed our attention on the combination of \gls{vae} and EEGNet. Specifically, we proposed \emph{vEEGNet}, a \gls{vae}-based \gls{dl} architecture consisting of two learning modules, i.e., an unsupervised representation learning module, and a supervised module, with the aim of both classifying \gls{mi} \gls{eeg} data and reconstructing them.
The first module was formed by a \gls{vae}~\cite{VAE_ORIGINAL_PAPER_kingma,MetroAgriFor2022,li2019disentangled}, while the second by an \gls{ffnn}.
In the \gls{vae}, we exploited EEGNet as the gls{vae} encoder (and, conversely, its mirrored version as the \gls{vae} decoder) to extract a compact and highly informative representation of the \gls{eeg} (with a latent space of dimension $d=16$). The latter was later used by the \gls{ffnn} to classify the \gls{eeg} into four different classes of movement. At the same time, that representation was used by the decoder to reconstruct the \gls{eeg} input data.
vEEGNet implemented a joint training of the two modules, by minimizing the joint loss function given by the sum of the \gls{vae} loss and the classifier loss.
Using such architecture, we were able to obtain classification performance at the state of the art and to reconstruct the low-frequency components of the input \gls{eeg} (in line with \cite{Bethge_2022_EEG2VEC}).

Therefore, in this work, we decided to extend our previous work by modifying the vEEGNet architecture (i) to create \emph{vEEGNet1}, i.e., very similar to our original vEEGNet but with a larger latent space ($d=64$), and (ii) to propose \emph{vEEGNet2}, which expands the original vEEGNet's encoder architecture to make it able to reconstruct not only the low frequency components of the input \gls{eeg}, but also the higher frequency ones (relevant for the studies on motor control)~\cite{Pfurtscheller2006}.

\section{Methods}
\label{sec:methods}

\subsection{Variational Autoencoder}
\label{subsec:VAE}

\textcolor{black}{The \gls{vae} is a variant of the standard auto-encoder model which takes advantage of an effective encoding-decoding \gls{dl} approach to obtain a projection of the data on a \emph{structured} latent space.
In fact, \gls{vae}s have been recently and successfully implemented for a number of unsupervised and semi-supervised learning problems, e.g., random sampling and interpolation~\cite{VAE_ORIGINAL_PAPER_kingma,hinton2006reducing,li2019disentangled,MetroAgriFor2022}.}

\textcolor{black}{In mathematical terms, the \gls{vae} is trained to learn two different probability distributions: a variational (approximate posterior) distribution $q_\phi(\mathbf{z} | \mathbf{x})$ of latent variables $\mathbf{z}$, given the observations $\mathbf{x}$ (i.e., at the encoder), as well as a generative model $p_\theta(\mathbf{x} | \mathbf{z})$ (i.e., at the decoder) \cite{blei2017variational}.}
This task is accomplished by the aforementioned pair of encoder-decoder networks, parameterized by $\phi$ and $\theta$, respectively.
\textcolor{black}{The \gls{vae} training aims to minimizing the \gls{vae} loss, $\mathcal{L}_{VAE}$, w.r.t. the parameters $\phi$ and $\theta$.}
$\mathcal{L}_{VAE}$ is usually expressed in term of the \gls{elbo} for the (evidence) probability $p(\mathbf{x})$, i.e., $\mathcal{L}(\theta, \phi; \mathbf{x})$: $\mathcal{L}_{VAE} = - \mathcal{L}(\theta, \phi; \mathbf{x})$, provided that

\begin{equation}
\mathcal{L}(\theta, \phi; \mathbf{x}) = \EX_{q_\phi(\mathbf{z} | \mathbf{x})}\left(log \frac{p_\theta(\mathbf{x},\mathbf{z})}{q_\phi(\mathbf{z} | \mathbf{x}))} \right).
\end{equation}


\noindent
Thus, for the \gls{vae} training, \textcolor{black}{the minimization of} $\mathcal{L}_{VAE}$ \textcolor{black}{leads to the maximization of} the \gls{elbo} for $p(\mathbf{x})$. 
The gap between $p(\mathbf{x})$ and $\mathcal{L}(\theta,\phi;\mathbf{x})$ can be best expressed by considering the Kullback-Leibler divergence ($\mathcal{KL}$) between the variational $q_\phi(\mathbf{z} | \mathbf{x})$ and posterior $p_\theta(\mathbf{x}|\mathbf{z})$ distributions, which turns to be
\begin{align}
    \mathcal{KL} [q_\phi(\mathbf{z}|\mathbf{x}) || p_\theta(\mathbf{x}|\mathbf{z})] = 
    - \mathcal{L}(\theta, \phi; \mathbf{x}) + p(\mathbf{x}) 
\end{align} 
Since $\mathcal{KL} [q_\phi(\mathbf{z}|\mathbf{x}) || p_\theta(\mathbf{x}|\mathbf{z})] \geq 0$, one arrives at the lower bound
$\mathcal{L}(\theta, \phi; \mathbf{x}) \leq p(\mathbf{x})$.
Similarly, the \gls{elbo} can be also formulated as
\begin{align}\label{eq:vae}
    \mathcal{L}(\mathbf{\theta}, \mathbf{\phi}; \mathbf{x}) & = \EX_q(p_\theta(\mathbf{x}|\mathbf{z})) - \mathcal{KL} [q_{\phi}(\mathbf{z}|\mathbf{x}) || p(\mathbf{z})] \\
    & = \mathcal{L}_{R} + \mathcal{L}_{KL}
\end{align}

\noindent
\textcolor{black}{In this way, the first term, $\mathcal{L}_{R} =\EX_q(p_\theta(\mathbf{x}|\mathbf{z}))$, can be interpreted as the likelihood of observing the original data given the input, and thus it provides the reconstruction error.
On the other hand, the second term, $\mathcal{L}_{KL} = -\mathcal{KL} [q_{\phi}(\mathbf{z}|\mathbf{x}) || p(\mathbf{z})]$, acts as a regularizer, thus penalizing those \emph{surrogate} distributions ($q_{\phi}(\mathbf{z}|\mathbf{x})$) too far away from the predefined $p(\mathbf{z})$.
}

\textcolor{black}{It is worth to note that the \gls{vae}s are able to provide output distributions that can be used to generate new samples.
In this work, we particularly exploits this opportunity by sampling samples from the latent space and using them to classify the different \gls{mi} movement classes. However, the investigation of the pure generation of new \gls{eeg} samples is beyond the scope of this article and it deserves a proper, separate, study.}

\begin{figure*}
    \centering
    \includegraphics[width = \linewidth]{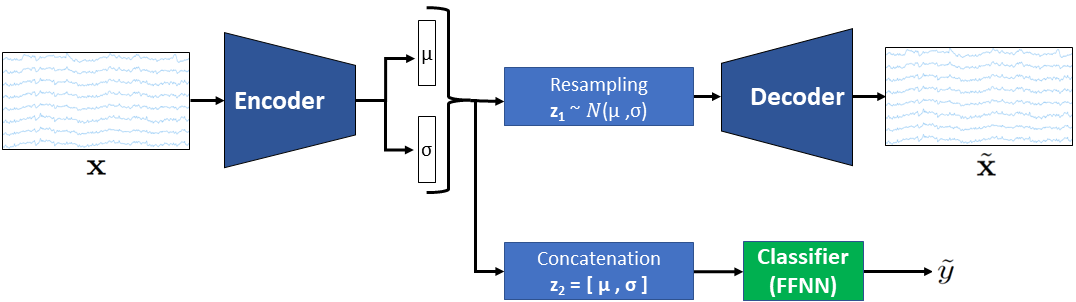}
    \caption{The general vEEGNet architecture (modified from our previous work \cite{Zancanaro_2023_vEEGNet}).}
    \label{fig:complete architecture}
\end{figure*}

\subsection{\textcolor{black}{vEEGNet general architecture}} \label{subsec:architecture}
\textcolor{black}{In this study, we developed a brand-new model which combines the advantages of \gls{vae}s~\cite{VAE_ORIGINAL_PAPER_kingma,li2019disentangled} and the specificity of EEGNet for classify and, at the same time, reconstruct \gls{eeg} raw data. In our previous works, we have already implemented these architectures for different studies, e.g., a $\beta$-\gls{vae} for anomaly detection in plants physiology~\cite{MetroAgriFor2022}, and an efficient version of the well-known EEGNet in~\cite{zancanaro_CIBCB_article}. However, here, we combine these two \gls{dl} models as shown in Fig.~\ref{fig:complete architecture}.}
\textcolor{black}{Particularly, we implemented EEGNet as one building block of the \gls{vae}, i.e., the encoder, and used it to extract a compact representation of the raw \gls{eeg} data, that can be further projected into a lower dimensional latent space. Then, the \gls{vae} decoder consisted of the \emph{mirrored} version of the same EEGNet architecture. To complete the architecture, we added an \gls{ffnn} which had the aim of classifying samples extracted from the latent space into one of the classes of interest.}
%
\textcolor{black}{Thus}, the model consists of two different mechanisms, ruled by an unsupervised and a supervised learning, respectively, as further explained \textcolor{black}{below}.

\subsubsection{Unsupervised mechanism}

The unsupervised mechanism exploits the EEGNet architecture to learn both the latent distribution 
$q_\phi(\mathbf{z} | \mathbf{x})$ as well as the posterior 
$p_\theta(\mathbf{x} | \mathbf{z})$. 
\textcolor{black}{For both the prior distribution $p(\mathbf{z})$ and the approximation posterior distribution $q_\phi(\mathbf{z} | \mathbf{x})$, we assumed isotropic Gaussian distribution, i.e.,}
\begin{align}
p(\mathbf{z}) & = \mathcal{N}(\mathbf{0}, \mathbf{I}) \\
q_{\phi}(\mathbf{z}|\mathbf{x}) & = \mathcal{N}(\mathbf{z};\mu(\mathbf{x};\phi), \sigma^2(\mathbf{x};\phi)\bm{I})
\end{align}
\noindent
where $\mu(\mathbf{x};\phi)$ and $\sigma(\mathbf{x};\phi)$ 
are the functions implemented by the vEEGNet encoder to encode the mean and the (diagonal) covariance matrix of the Gaussian distribution. With these assumptions, $\mathcal{KL} [q_{\phi}(\mathbf{z}|\mathbf{x}) || p(\mathbf{z})]$ (the regularization term defined in Section \ref{subsec:VAE}) can be directly expressed in the compact analytical form:
\begin{align}
\mathcal{L}_{KL} & = \mathcal{KL} [q_{\phi}(\mathbf{z}|\mathbf{x}) || p(\mathbf{z})] = \frac{1}{2}\sum_{i = 1}^{d}(\sigma^2_i+ \mu^2_i - 1 - log(\sigma^2_i))
\end{align}
\noindent
where $\mu_i$ and $\sigma^2_i$ are the predicted mean and variance values of the corresponding $i$-th latent component of $\mathbf{z}$~\cite{VAE_ORIGINAL_PAPER_kingma}.
%
The vEEGNet encoder implements a standard EEGNet with its usual blocks, i.e., a temporal convolution, a spatial convolution, and a separable convolution. Lastly, the output is flattened and given to a fully-connected layer.
\textcolor{black}{
From the vEEGNet encoder's output (i.e., giving $q_{\phi}(\mathbf{z}|\mathbf{x})$), we sampled a vector, say $\mathbf{z}_1$ (using the reparametrization trick $\mathbf{z}_{1}  = \mu + \sigma \cdot N(\mathbf{0}, \bm{1})$). Then, $\mathbf{z}_1$ is input to the vEEGNet decoder which had the ultimate goal to reconstruct the original raw \gls{eeg} signal.} 
\textcolor{black}{The decoder architecture mirrors that of the encoder, i.e., with a sequence of layers made by a separable convolution, a spatial filter, a temporal filter. Transposed convolutions are used instead of normal convolutions, and up-sample layers in place of standard pooling layers.}
In both the encoder and the decoder, batch normalization and dropout layers were added to increase performance and stability during training.
\textcolor{black}{To compute the reconstruction loss during the training, we computed $\mathcal{L}_{R}$ through the \gls{mse}.}

\subsubsection{Supervised mechanism}
The supervised mechanism is given by the \gls{ffnn} \textcolor{black}{which aimed at classifying the raw \gls{eeg}} into $n$ different classes, \textcolor{black}{with $n$ the specific number of classes given by the dataset}.
In vEEGNet, a second vector $\mathbf{z}_2 = [\bm{\mu}, \bm{\sigma}^2]$ is obtained by concatenating the output of the encoder, i.e., the parameters vectors $\Tilde{\mu}=\mu(\mathbf{x};\bm{\phi})$ and $\Tilde{\sigma} = \sigma(\mathbf{x};\bm{\phi})$. 
This new vector is fed into the classifier to output the predicted class $\Tilde{y}$. 
For the classifier, we used the negative log-likelihood loss function defined as:
\begin{equation}
    \mathcal{L}_{clf} = - \log (\Tilde{\mathbf{y}}) \cdot \mathbf{y}
\end{equation}
where $\log (\Tilde{\mathbf{y}})$ are the log probabilities of the possible labels related to input $\mathbf{x}$, and $\mathbf{y}$ is a one hot encoded vector of the true labels of input $\mathbf{x}$.

\textcolor{black}{Finally, vEEGNet aimed to minimize the overall loss} function $\mathcal{L}_{Total}$ given by the sum of the \gls{vae} loss and the classifier loss ($\mathcal{L}_{clf}$), as follows: 
\begin{align}
    \begin{split}
        \mathcal{L}_{Total} & = \mathcal{L}_{VAE} + \mathcal{L}_{clf}
         = \mathcal{L}_{R} + \mathcal{L}_{KL} + \mathcal{L}_{clf}
    \end{split}
\end{align}

\section{Results and discussion}
\label{sec:results}

\subsection{Dataset and vEEGNet implementations} \label{subsec:dataset}
\textcolor{black}{We applied vEEGNet to the public \emph{dataset 2a} of the IV BCI competition~\cite{dataset_BCI_competition}. The dataset includes 22-channel \gls{eeg} recordings from 9 subjects performing imagination of 4 different movements, i.e., the \gls{mi} of either right or left hand, both feet or tongue.}
\textcolor{black}{The training set consists of 288 trials (or repetitions) for each subject, while the test set consists of 288 different trials for each subject.} The \gls{eeg} data have been previously filtered with a $0.5-100\SI{}{\hertz}$ band-pass filter and a notch filter at 50 Hz.
\textcolor{black}{We down-sampled the data to $\SI{128}{\hertz}$, following the same approach of other works~\cite{MI_EEGNet_AC_TAB_10,EEGNet_paper}, including our own~\cite{zancanaro_CIBCB_article}.} Then, for every \gls{mi} repetition, \textcolor{black}{a $\SI{4}{\second}$ \gls{eeg} segment was extracted (channel-wise)}, thus obtaining a $22 \times 512$ data matrix.

\begin{figure*}
    \centering
        \begin{subfigure}[b]{1\textwidth}
        \includegraphics[width = \linewidth]{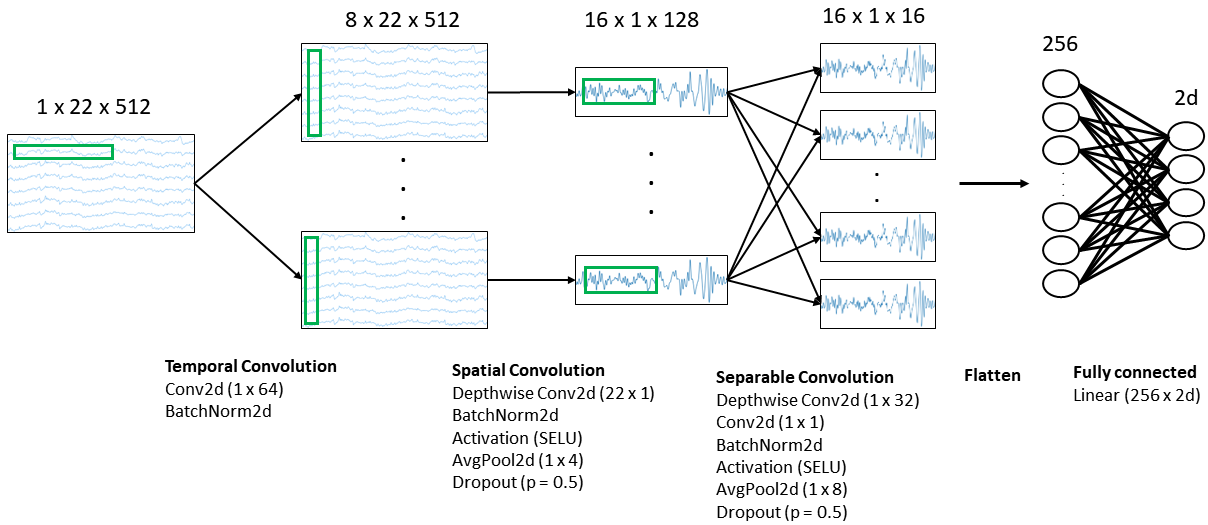}
        \caption{Encoder.}
        \label{fig:vEEGNet12_encoder}
    \end{subfigure}

\vspace{1cm}
    
    \begin{subfigure}[b]{1\textwidth}
        \includegraphics[width = \linewidth]{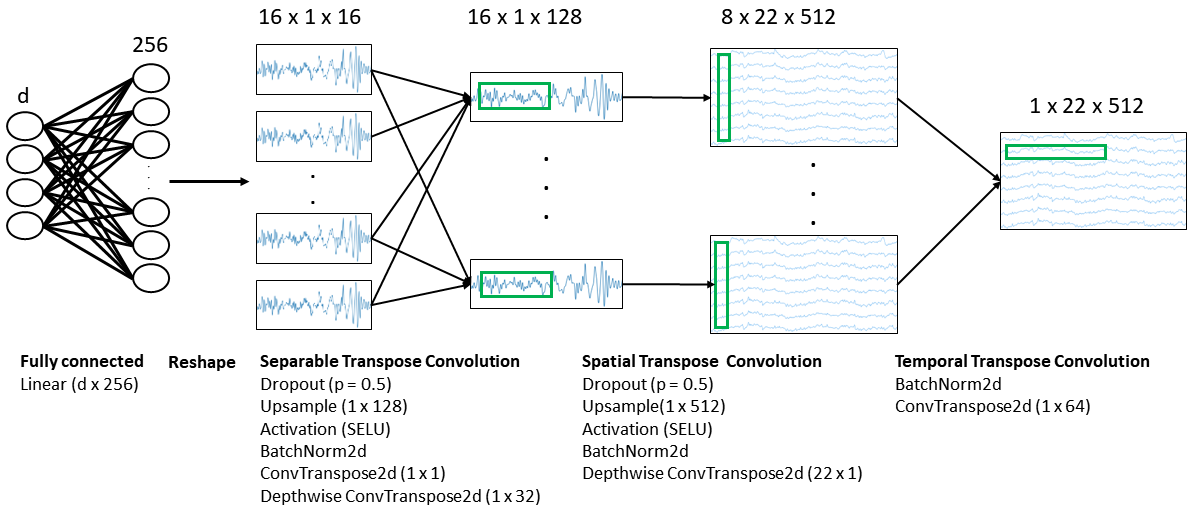}
        \caption{Decoder}
        \label{fig:vEEGNet12_decoder}
    \end{subfigure}
    \caption{\textcolor{black}{The vEEGNet1 architecture, with all implementation details (i.e., parameters and activation functions).}}
    \label{fig:vEEGNet12}
\end{figure*}

\textcolor{black}{We implemented vEEGNet in PyTorch \footnote{\tiny{The code is available on GitHub: https://github.com/jesus-333/Variational-Autoencoder-for-EEG-analysis/tree/code\_ICT4AWE\_2023}} and we trained it using RTX 2070, 500 epochs, AdamW optimizer~\cite{AdamW_paper}, a learning rate of $0.001$, and a weight decay of $0.00001$.
We implemented two slightly, but significantly different, architectures, namely \emph{vEEGNet1} and \emph{vEEGNet2}. They differ in the encoder and the classification network.
%
In vEEGNet1, we simply implemented the original EEGNet \cite{EEGNet_paper} as the encoder and its mirrored version as the decoder with a latent space of dimension $d=64$. The full architecture is shown in detail in Fig.~\ref{fig:vEEGNet12}. Moreover, the classifier consists of an input layer with 128 neurons, followed by one hidden layer with 64 neurons, and an ELU activation function. The output layer has 4 neurons to classify the 4 different \gls{mi} tasks and implements a log-softmax activation function.
%
On the other hand, in vEEGNet2, we further expanded the encoder and implemented it as 3 parallel EEGNet networks (i.e., 3 sub-encoders), as inspired by the work of \cite{MBShallow_CovNet_AC_TAB_10}. The decoder remained unchanged.
Interestingly, the 3 sub-encoders differed from each other by the length of the kernels in the first layer, i.e., the layer responsible for time convolutions. In fact, using temporal kernels with different lengths makes it possible to specialize each sub-encoder in learning features with different frequency contents (i.e., different frequency bands). Therefore, we selected the values of 64, 16 and 4, respectively, for the kernel lengths of the 3 sub-encoders. Finally, the vEEGNet2 classifier consists of an input layer with 128 neurons, followed by an output layer with 4 neurons (with the log-softmax activation function kept to obtain the classification and no hidden layers).
To note, we trained vEEGNet2 using the min-max normalized \gls{eeg} data, i.e., with values in the range $[-1, 1]$, as we observed an improvement of its reconstruction ability.}

%
The total number of trainable parameters for vEEGNet1 is $61476$, with $52960$ of them for the implementation of the unsupervised mechanism and the remaining $8516$ for the supervised one.
\textcolor{black}{On the other hand, the total number of trainable parameters for vEEGNet2 is $121853$, with $516$ parameters belonging to the architecture part that implements the classifier. Note that the decreased number of parameters for vEEGNet2 classifier is due to the absence of any hidden layer in the classifier.}
\textcolor{black}{To note, for both implementations, from our previous empirical evaluations \cite{Zancanaro_2023_vEEGNet}, we chose $d = 64$ as the hidden space dimension for the latent space extracted by the encoder.
In line with a common empirical approach, and with previous works, including ours~\cite{Wang_2020_EEGNet_autoencoder_1,Zancanaro_2023_vEEGNet}, we considered the first $d/2$ neurons as the vector of the means ($\bm{\mu}$), and the remaining $d/2$ neurons as the vector of the variances ($\bm{\sigma^2}$).}
%
%

\subsection{vEEGNet performance}
Both vEEGNet1 and vEEGNet2 were evaluated for their classification and reconstruction abilities on the \emph{dataset 2a}. Particularly, vEEGNet1 was able to classify the 4 different imagined movements with a performance in line with the most recent state of the art.
Table~\ref{tab:Accuracy comparison V2} reports the classification accuracy and the Cohen's $\kappa$ score obtained by vEEGNet1 for every subject. The performance of other relevant and recent \gls{dl} models  are reported as a comparison, including the authors' previous EEGNet implementation through \emph{DynamicNet} (a tool to facilitate the implementation of neural networks~\cite{zancanaro_CIBCB_article}).
Only models classifying all 4 types of movements were included in this comparison (i.e., with a chance level of $0.25$).
From these results, it is possible to notice how the inter-subject inherent variability influenced the classification of the different models.
%

%


%

\begin{table*}
\centering
\caption{ Comparison of \textcolor{black}{vEEGNet1} with other \gls{dl} models in terms of classification accuracy ($[\%]$) and kappa score (when available, its value is within brackets) in a \textcolor{black}{4-class} \gls{mi} task. The first five columns refer to EEGNet-based models, while the last three columns refer to general-purpose \gls{cnn} models. AVG stands for average, STD for standard deviation. \textcolor{black}{The table was slightly modified from our original conference paper \cite{Zancanaro_2023_vEEGNet}.}}
\label{tab:Accuracy comparison V2}
\resizebox{\textwidth}{!}{%
\begin{tabular}{|c|c||c|c|c|c|c|c|c|} 
\cline{2-9}
\multicolumn{1}{c|}{} 
&
\begin{tabular}[c]{@{}c@{}}\textbf{\textcolor{black}{vEEGNet1}}\\\textbf{(\textcolor{black}{d = 64})}\end{tabular}
&
\begin{tabular}[c]{@{}c@{}}\textbf{EEGNet}
\end{tabular}
&
\begin{tabular}[c]{@{}c@{}}\textbf{TSGL-}\\\textbf{EEGNet}
\end{tabular}
& \textbf{MI-EEGNet}
& \begin{tabular}[c]{@{}c@{}}\textbf{MBShallow}\\\textbf{ConvNet~}
\end{tabular} 
& \begin{tabular}[c]{@{}c@{}}\textbf{CW-CNN} 
\end{tabular} 
& \begin{tabular}[c]{@{}c@{}}\textbf{DFFN}
\end{tabular} 
& 
\begin{tabular}[c]{@{}c@{}}\textbf{Monolithic}\\\textbf{Network}
\end{tabular}  \\ 
\hline
\textbf{1}    & 70.83   & 81.88   & 85.41~ (0.81)  & 83.68~ (0.78)   & 82.58~ (0.77)                          & 86.11~ (0.82)  & 83.2    & 83.13~ (0.67)                                                                  \\
\textbf{2}            & 57.64                                                               & 60.97                                                                                                    & 70.67~ (0.61)                                                                      & 49.65~ (0.33)          & 70.01~ (0.6)                                                                 & 60.76~ (0.48)                                                      & 65.69                                                            & 65.45~ (0.35)                                                                  \\
\textbf{3}            & 85.76                                                               & 88.54                                                                                                    & 95.24~ (0.94)                                                                      & 89.24~ (0.86)          & 93.79~ (0.92)                                                                & 86.81~ (0.82)                                                      & 90.29                                                            & 80.29~ (0.65)                                                                  \\
\textbf{4}            & 61.46                                                               & 70.63                                                                                                    & 80.26~ (0.74)                                                                      & 68.06~ (0.57)          & 82.6~ (0.77)                                                                 & 67.36~ (0.57)                                                      & 69.42                                                            & 81.6~ (0.62)                                                                   \\
\textbf{5}            & 60.42                                                               & 68.45                                                                                                    & 70.29~ (0.6)                                                                       & 64.93~ (0.53)          & 77.81~ (0.7)                                                                 & 62.5~ (0.5)                                                        & 61.65                                                            & 76.7~ (0.58)                                                                   \\
\textbf{6}            & 62.85                                                               & 61.46                                                                                                    & 68.37~ (0.58)                                                                      & 56.25~ (0.42)          & 64.79~ (0.53)                                                                & 45.14~ (0.27)                                                      & 60.74                                                            & 71.12~ (0.45)                                                                  \\
\textbf{7}            & 71.88                                                                & 82.08                                                                                                    & 90.97~ (0.88)                                                                      & 94.1~ (0.92)           & 88.02~ (0.84)                                                                & 90.63~ (0.88)                                                      & 85.18                                                            & 84~ (0.69)                                                                     \\
\textbf{8}            & 77.43                                                               & 82.15                                                                                                    & 86.35~ (0.82)                                                                      & 82.64~ (0.77)          & 86.91~ (0.83)                                                                & 81.25~ (0.75)                                                      & 84.21                                                            & 82.66~ (0.7)                                                                   \\
\textbf{9}            & 65.28                                                               & 66.25                                                                                                    & 83.64~ (0.79)                                                                      & 82.99~ (0.77)          & 83.38~ (0.78)                                                                & 77.08~ (0.69)                                                      & 85.48                                                            & 80.74~ (0.64)                                                                  \\ 
\hline
\textbf{AVG}          & \textbf{68.17}                                                      & \textbf{73.60}                                                                                           & \textbf{81.34~ (0.75)}                                                             & \textbf{74.61~ (0.66)} & \textbf{81.15~ (0.75)}                                                       & \textbf{73.07~ (0.64)}                                             & \textbf{76.44}                                                   & \textbf{78.1~ (0.59)}                                                          \\ 
\hline
\textbf{STD}          & \textbf{9.14}                                                       & \textbf{10.20}                                                                                           & \textbf{9.61~ (0.13)}                                                              & \textbf{15.44~ (0.21)} & \textbf{9.03~ (0.12)}                                                        & \textbf{15.11~ (0.2)}                                              & \textbf{11.65}                                                   & \textbf{6.27~ (0.12)}                                                          \\
\hline
\end{tabular}}
\end{table*}


%


\textcolor{black}{As from the classification results reported in Table~\ref{tab:Accuracy comparison V2}, we observed that
those models which combine multiple EEGNet \emph{units} (e.g., TSGL-EEGNet, MBShallow ConvNet) can reach higher performance (up to $80\%$, with improvements of 2-8\% w.r.t. to the other models), we implemented vEEGNet2.
Unfortunately, vEEGNet2 did not improve the classification performance as expected, e.g., barely reaching a maximum accuracy value for $64.58$ for subject $1$.}





\begin{figure} 
\centering
    \begin{subfigure}[b]{0.49\textwidth}
        \includegraphics[height = 4cm, width=\textwidth]{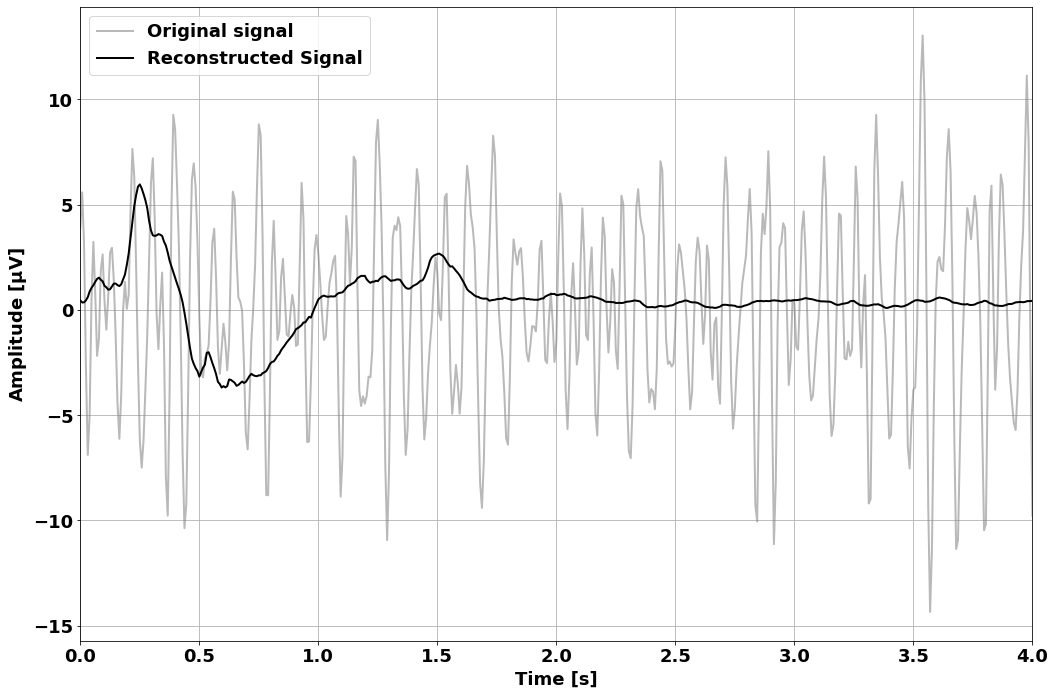}
        \caption{vEEGNet1}
        \label{fig:reconstructed_EEG_vEEGNet1}
    \end{subfigure}
    \begin{subfigure}[b]{0.49\textwidth}
        \includegraphics[height = 4cm, width=\textwidth]{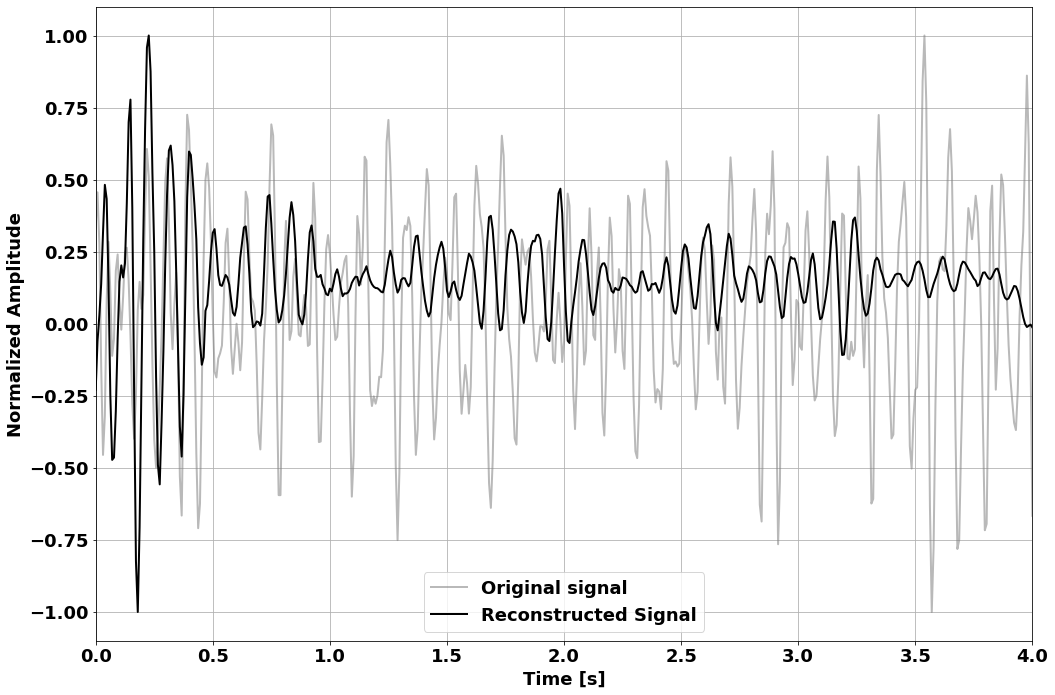}
        \caption{vEEGNet2}
        \label{fig:reconstructed_EEG_vEEGNet2}
    \end{subfigure}

\caption{An example of \textcolor{black}{an \gls{eeg} signal (channel C3, right-hand MI) as reconstructed by (a) vEEGNet1 and (b) vEEGNet2.}}
\label{fig:reconstructed_EEG}
\end{figure}
\begin{figure} 
    \centering
    \begin{subfigure}[b]{0.49\textwidth}
        \includegraphics[height = 4cm, width=\textwidth]{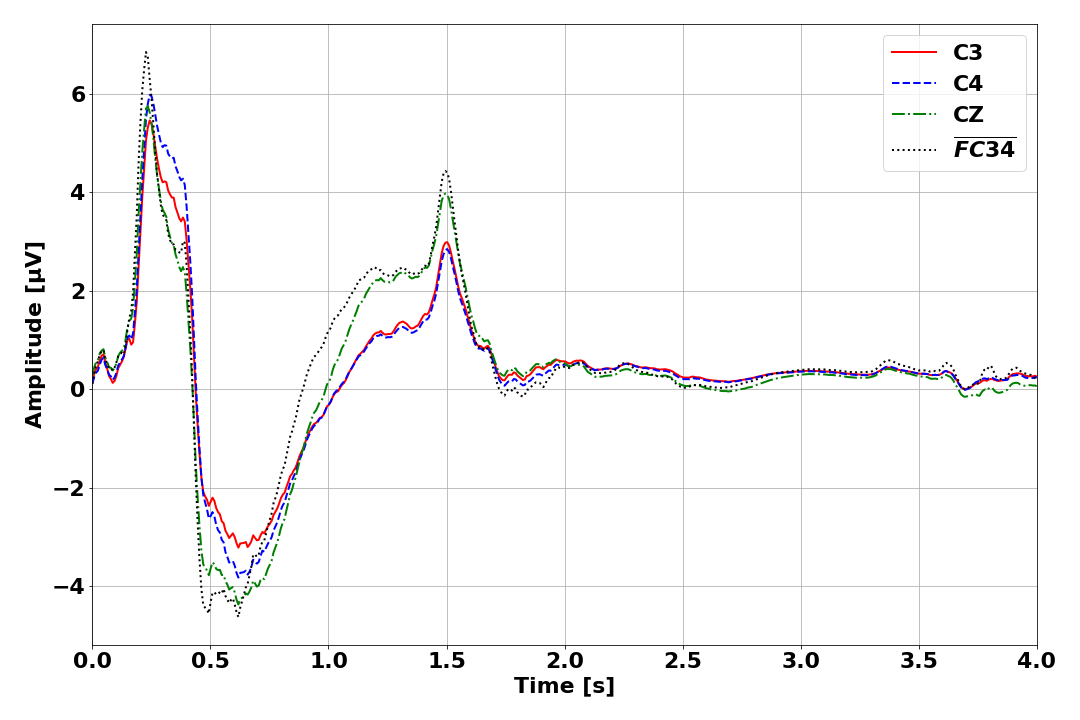}
        \caption{vEEGNet}
        \label{fig:MRCP_shape_vEEGNet_1}
    \end{subfigure}
    \begin{subfigure}[b]{0.49\textwidth}
         \includegraphics[height = 4cm, width=\textwidth]{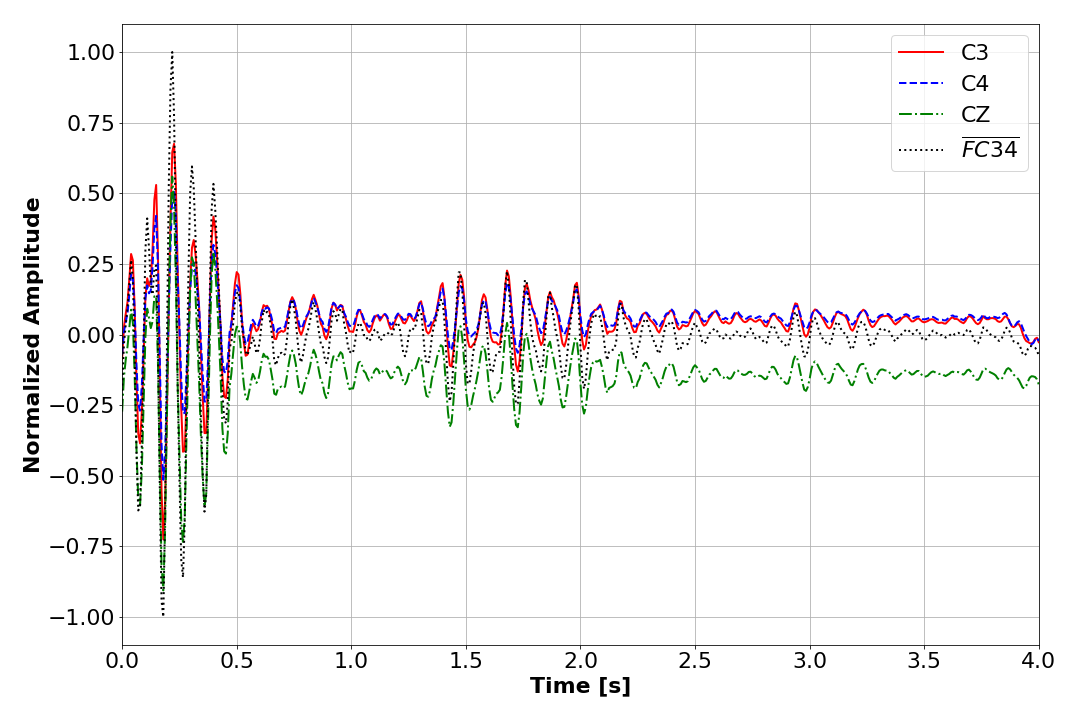}
        \caption{vEEGNet2}
        \label{fig:MRCP_shape_vEEGNet_2}
    \end{subfigure}
    \caption{\textcolor{black}{An example of 4 different \gls{eeg} signals from 4 target sensors (namely, C3, C4, Cz, and the average between FC3 and FC4) as reconstructed by (a) vEEGNet and (b) vEEGNet2. Panel (a) has been reported from our previous conference paper \cite{Zancanaro_2023_vEEGNet} to show the difference with the results from vEEGNet2.}}
    \label{fig:MRCP_shape}
\end{figure}

\textcolor{black}{Nevertheless, as the reconstruction task as regards, both vEEGNet1 and vEEGNet2 showed their own potentialities and specificity.}
\textcolor{black}{Fig.~\ref{fig:reconstructed_EEG} reports an example of \gls{eeg} signal (channel C3, right-hand \gls{mi}) as reconstructed by vEEGNet1 (Fig.~\ref{fig:reconstructed_EEG}(a)) and vEEGNet2 (Fig.~\ref{fig:reconstructed_EEG}(b)), respectively.}
%
In fact, from the reconstruction obtained by vEEGNet1, we recognize the so-called \gls{mrcp}, a specific and well-known \gls{eeg} component that typically appears when a movement is executed or imagined. \gls{mrcp}s are low-frequency components (in the range of $0.5$-$4$~Hz) that are characterized by a sequence of a positive peak occurring right after the beginning of the movement, a negative peak (within $1$~s), and finally a rebound to positive values. The \gls{mrcp} tends to expire (i.e., return to baseline values) within approximately $2$~s after the beginning of the movement. Therefore, in Fig.~\ref{fig:reconstructed_EEG} (a), we could observe a pattern that is very consistent with the one expected by the literature~\cite{Magnuson2021,li2018,Bressan2021}.
%
\textcolor{black}{On the other hand, vEEGNet2 showed the ability to reconstruct middle-range frequency components (approximately in the range $5$-$20$~Hz).}
\textcolor{black}{Both frequency bands, i.e., the low and the middle range, have a critical importance in neuroscience studies related to movement \cite{Bressan2021} and, as such, both vEEGNet1 and vEEGNet2 have their own potentialities to be useful in this context.}
%
%
%
Fig.~\ref{fig:MRCP_shape} shows four different reconstructed \gls{eeg} channels, namely C3, C4, Cz, and the average of FC3 and FC4, selected based on their relevance to the \gls{mi} tasks. In line with well-known literature~\cite{Lazurenko2018}, the most relevant electrodes where to retrieve information related to the hand movement (i.e., target sensors) are C3 and C4, for the movements of the right and the left hand, respectively, Cz for the feet, and sensors F3 and F4 for the tongue. Since, F3 and F4 were not available in this dataset, then we considered the nearest available sensors which were FC3 and FC4 (based on the International $10$-$20$ System for \gls{eeg} electrode placement).As expected from the literature \cite{SeelandMRCP2015}, all reconstructed signals from the target sensors displayed the \gls{mrcp} and are suitable for a further neurophysiological investigation.
%
%
However, it is possible to note that both architectures are not fully able to reconstruct the entire 4s \gls{eeg} segments. Specifically, they can follow the raw data behaviour with high-fidelity in the initial part (actually, very critical to study movement initiation \cite{bodda2022,kobler2020}), but then the reconstructed signals vanish. Also, the output from vEEGNet2 has been magnified to the range of $[-1,1]$ to have a scale that is comparable with the raw data (i.e., scaled in the same range as explained above).
These issues represent limitations of our architectures that deserve further investigations. However, other works have recently showed similar problems in reconstructing \gls{eeg} signals. As an example, in \cite{Bethge_2022_EEG2VEC}, the authors used an EEGNet-based \gls{vae} to encode and then reconstruct the \gls{eeg} data related to emotions. They succeeded in reconstructing only a low-frequency signal. In addition, the amplitude of the reconstructed signal was about half that of the original one.
\cite{Aznan_2019_eeg_generation} used a \gls{vae} to generate steady-state visual evocked potentials (SSVEP)-related \gls{eeg} data at 3 different frequencies, i.e., \SI{10}{\hertz}, \SI{12}{\hertz}, and \SI{15}{\hertz}. The reconstructed signals resembled the original ones but theirs amplitude was much smaller.



%
%
%
%

%

%
%


%
%
\begin{table}
\centering
\caption{Reconstruction error for vEEGNet1}
\label{tab:recon_vEEGNet_1}
\resizebox{0.8\linewidth}{!}{%
\begin{tblr}{
  width = \linewidth,
  colspec = {Q[213]Q[169]Q[169]Q[169]Q[169]},
  cells = {c},
  cell{1}{2} = {c=2}{0.338\linewidth},
  cell{1}{4} = {c=2}{0.338\linewidth},
  cell{12}{2} = {c=2}{0.338\linewidth},
  cell{12}{4} = {c=2}{0.338\linewidth},
  cell{13}{2} = {c=2}{0.338\linewidth},
  cell{13}{4} = {c=2}{0.338\linewidth},
    vline{2-3,5} = {-}{},
  vline{4,6} = {-}{},
  vline{1-3,4,6} = {2-11}{},
  vline{1-3,5} = {12-13}{},
  hline{1} = {2-5}{},
  hline{2-3,12,14} = {-}{},
}
        & Train set &       & Test set &       \\
Subject & AVG       & STD   & AVG      & STD   \\
1       & 27.82     & 5.79  & 28.22    & 5.75  \\
2       & 29.30     & 6.20  & 66.13    & 54.88 \\
3       & 50.16     & 15.37 & 47.55    & 13.70 \\
4       & 24.63     & 4.62  & 26.31    & 6.01  \\
5       & 17.23     & 3.26  & 21.71    & 4.10  \\
6       & 48.46     & 13.12 & 44.04    & 13.07 \\
7       & 22.33     & 4.68  & 27.71    & 5.95  \\
8       & 80.76     & 22.17 & 70.49    & 15.32 \\
9       & 84.19     & 23.64 & 84.72    & 27.86 \\
AVG     & 42.76  &       & 46.32 &       \\
STD     & 10.98  &       & 16.29 &   
\end{tblr}
}
\end{table}

\textcolor{black}{
To further evaluate the reconstruction capability of both vEEGNet1 and vEEGNet2, we also report the average reconstruction error for the two different implementations.
Tab.~\ref{tab:recon_vEEGNet_1} and Tab.~\ref{tab:recon_vEEGNet_2} show the average (across segments) of the \gls{mse} computed between every raw 4s \gls{eeg} segment and its reconstructed version, subject-wise. The values are independently reported for the training and the test set, respectively.
As expected, it can be observed (e.g., Tab.~\ref{tab:recon_vEEGNet_1}) the high variability of the reconstruction quality for different subjects (with \gls{mse} values ranging between $17.23$ and $84.19$ in the training set and from $21.71$ to $84.72$ in the test set). Also, we can further notice that some subjects have a higher degree of inherent intra-subject variability, with large standard deviations compared to the average (e.g., subjects $2$ in the test set). This might have increased the difficulty in reconstructing high-quality \gls{eeg} signals.
Incidentally, we might speculate that in cases such that of subject 2, i.e., when there is a huge difference between the test and training set results, overfitting might have been occurred. Alternatively, the high standard deviation might equally be due to the presence of anomalies (due to the recording setup as well as the physiological condition of the subject) that greatly increase the variability of the \gls{eeg} data and, consequently, make it more difficult to accurately reconstruct the signals.
Despite the different ranges for the average \gls{mse} values (due to the normalization operated in vEEGNet2), we can see from Tab.~\ref{tab:recon_vEEGNet_2} that the results obtained from vEEGNet2 are far more robust. The \gls{eeg} of all subjects can be reconstructed with an average \gls{mse} which is about $0.11$ with a small standard deviation (about one order of magnitude smaller than the corresponding average value). Moreover, the performance in the test set are rather comparable with the training set, with no cases of significant drop of the reconstruction quality.
}


\begin{table}
\centering
\caption{Reconstruction error for vEEGNet2}
\label{tab:recon_vEEGNet_2}
\resizebox{0.8\linewidth}{!}{%
\begin{tblr}{
  width = \linewidth,
  colspec = {Q[213]Q[169]Q[169]Q[169]Q[169]},
  cells = {c},
  cell{1}{2} = {c=2}{0.338\linewidth},
  cell{1}{4} = {c=2}{0.338\linewidth},
  cell{12}{2} = {c=2}{0.338\linewidth},
  cell{12}{4} = {c=2}{0.338\linewidth},
  cell{13}{2} = {c=2}{0.338\linewidth},
  cell{13}{4} = {c=2}{0.338\linewidth},
    vline{2-3,5} = {-}{},
  vline{4,6} = {-}{},
  vline{1-3,4,6} = {2-11}{},
  vline{1-3,5} = {12-13}{},
  hline{1} = {2-5}{},
  hline{2-3,12,14} = {-}{},
}
        & Train set &       & Test set &       \\
Subject & AVG       & STD   & AVG      & STD   \\
1       & 0.112     & 0.013 & 0.115    & 0.012 \\
2       & 0.111     & 0.015 & 0.118    & 0.015 \\
3       & 0.113     & 0.015 & 0.117    & 0.015 \\
4       & 0.111     & 0.013 & 0.111    & 0.013 \\
5       & 0.097     & 0.016 & 0.097    & 0.015 \\
6       & 0.112     & 0.012 & 0.112    & 0.013 \\
7       & 0.108     & 0.012 & 0.107    & 0.013 \\
8       & 0.117     & 0.013 & 0.115    & 0.012 \\
9       & 0.118     & 0.014 & 0.122    & 0.013 \\
AVG     & 0.111  &       & 0.113 &       \\
STD     & 0.006  &       & 0.007 &   
\end{tblr}
}
\end{table}


%
%

%
%
%


\section{Conclusions}
\label{sec:conclusions}

\textcolor{black}{
In this work, we tackled the challenging task of developing a \gls{dl} model which is both able to classify different types of \gls{eeg} data, e.g., different types of imagined movements (\gls{mi}), and to reconstruct the raw \gls{eeg} signals with high fidelity. Providing a solution to these challenges would have such a great impact in many different applications, e.g., from denoising to data generation in neuroscience and \gls{bci}.
}

\textcolor{black}{
After reviewing the related work, we directed our efforts on developing an architecture which combines the reconstructing and generative potentialities of the \gls{vae} models with the well-known \gls{cnn}-based EEGNet model, which has been specifically created to process \gls{eeg} data, namely vEEGNet, which have preliminarily been proposed in our previous conference paper. 
}

\textcolor{black}{
In this work, we presented two different implementations of vEEGNet, namely vEEGNet1 and vEEGNet2, and we applied them to a well-known public 22-channel \gls{eeg} dataset, i.e., the \emph{dataset 2a} from the IV BCI competition, to both classify 4 different types of imagined movements and reconstruct them.
We showed that vEEGNet1 is able to reach state of the art accuracy values in the classification task (around $70\%$ w.r.t. a chance level of $25\%$) and to reconstruct a low-frequency component from the input data. Interestingly, we have recognized such a component as the well-known physiological phenomenon called \gls{mrcp}.
To improve the reconstruction performance and inspired by recent literature, we further expanded vEEGNet into vEEGNet2, which implemented 3 parallel EEGNet networks to form its encoder. Thus, with this design choice, we were able to capture faster \gls{eeg} components and reconstruct them. With vEEGNet2, we were able to significantly overcome the limitations of other previous works \cite{Bethge_2022_EEG2VEC,Zancanaro_2023_vEEGNet} which proposed \gls{dl} architectures that could reconstruct low-frequency components, only.
%
}

\textcolor{black}{
Overall, vEEGNet in its two implementations achieved good classification results and promising reconstruction performance, improving the state of the art. However, this contribution is still preliminary and, as such, a number of limitations and open challenges are also discussed, and will need further investigations.
Among others, from this work we found out that achieving a good classification accuracy does not necessarily means to have a good quality reconstruction, i.e., as observed from the classification results of vEEGNet2. However, this is in line with other works, e.g., the EEG2VEC \cite{Bethge_2022_EEG2VEC} showed a very good classification performance, at the expenses of a rather limited quality in the reconstruction.
Furthermore, future investigations might be also directed to explore the potentialities of vEEGNet \textcolor{black}{(and its different implementations)} as generative model for \gls{eeg} in order to cope with the common lack of large \gls{eeg} datasets thus helping \gls{dl} models to improve their performance and better generalize over different \gls{eeg} datasets.
}

\subsection*{Acknowledgements}
This work was partially supported by the MUR under the grant “Dipartimenti di Eccellenza 2023-2027” of the Department of Informatics, Systems and Communication of the University of Milano-Bicocca, Italy.

AZ is also supported by PON 2014-2020 action IV.4 funded by the Italian Ministry of University and Research at the University of Padova (Padova, Italy).

GC is also supported by PON Initiative 2014-2020 action IV.6  funded by the Italian Ministry of University and Research at the University of Milan-Bicocca (Milan, Italy).

%
%
\bibliographystyle{splncs04}
\bibliography{0_main}

\begin{thebibliography}{10}
\providecommand{\url}[1]{\texttt{#1}}
\providecommand{\urlprefix}{URL }
\providecommand{\doi}[1]{https://doi.org/#1}

\bibitem{MBShallow_CovNet_AC_TAB_10}
Altuwaijri, G.A., Muhammad, G.: A multibranch of convolutional neural network models for electroencephalogram-based motor imagery classification. Biosensors  \textbf{12}(1) (2022). \doi{10.3390/bios12010022}, \url{https://www.mdpi.com/2079-6374/12/1/22}

\bibitem{yang_2018_DSAE_EEG_dataset}
Andrzejak, R., Lehnertz, K., Mormann, F., Rieke, C., David, P., Elger, C.: Indications of nonlinear deterministic and finite-dimensional structures in time series of brain electrical activity: Dependence on recording region and brain state. Physical review. E, Statistical, nonlinear, and soft matter physics  \textbf{64},  061907 (01 2002). \doi{10.1103/PhysRevE.64.061907}

\bibitem{Aznan_2019_eeg_generation}
Aznan, N., Atapour~Abarghouei, A., Bonner, S., Connolly, J., Al~Moubayed, N., Breckon, T.: Simulating brain signals: Creating synthetic eeg data via neural-based generative models for improved ssvep classification. pp.~1--8 (07 2019). \doi{10.1109/IJCNN.2019.8852227}

\bibitem{Bethge_2022_EEG2VEC}
Bethge, D., Hallgarten, P., Grosse-Puppendahl, T., Kari, M., Chuang, L.L., Özdenizci, O., Schmidt, A.: Eeg2vec: Learning affective eeg representations via variational autoencoders. In: 2022 IEEE International Conference on Systems, Man, and Cybernetics (SMC). pp. 3150--3157 (2022). \doi{10.1109/SMC53654.2022.9945517}

\bibitem{dataset_BCI_competition}
Blankertz, B., Dornhege, G., Krauledat, M., Müller, K.R., Curio, G.: The non-invasive {B}erlin {B}rain-{C}omputer {I}nterface: Fast acquisition of effective performance in untrained subjects. NeuroImage  \textbf{37},  539--50 (09 2007). \doi{10.1016/j.neuroimage.2007.01.051}

\bibitem{blei2017variational}
Blei, D.M., Kucukelbir, A., McAuliffe, J.D.: Variational inference: A review for statisticians. Journal of the American statistical Association  \textbf{112}(518),  859--877 (2017)

\bibitem{bodda2022}
Bodda, S., Diwakar, S.: Exploring eeg spectral and temporal dynamics underlying a hand grasp movement. Plos one  \textbf{17}(6),  e0270366 (2022)

\bibitem{Bressan2021}
Bressan, G., Cisotto, G., M{\"u}ller-Putz, G.R., Wriessnegger, S.C.: Deep learning-based classification of fine hand movements from low frequency {eeg}. Future Internet  \textbf{13}(5), ~103 (2021)

\bibitem{ICC2020}
Cisotto, G., Capuzzo, M., Guglielmi, A.V., Zanella, A.: Feature selection for gesture recognition in internet-of-things for healthcare. In: ICC 2020-2020 IEEE International Conference on Communications (ICC). pp.~1--6. IEEE (2020)

\bibitem{EURASIP2022}
Cisotto, G., Capuzzo, M., Guglielmi, A.V., Zanella, A.: Feature stability and setup minimization for {EEG}-{EMG}-enabled monitoring systems. EURASIP Journal on Advances in Signal Processing  \textbf{2022}(1), ~103 (2022)

\bibitem{ICC2013}
Cisotto, G., Pupolin, S., Silvoni, S., Cavinato, M., Agostini, M., Piccione, F.: Brain-computer interface in chronic stroke: An application of sensorimotor closed-loop and contingent force feedback. In: 2013 IEEE International Conference on Communications (ICC). pp. 4379--4383. IEEE (2013)

\bibitem{TSGL_EEGNet_AC_TAB_9}
Deng, X., Zhang, B., Yu, N., Liu, K., Sun, K.: Advanced {TSGL-EEGN}et for motor imagery {EEG}-based {B}rain-{C}omputer {I}nterfaces. IEEE Access  \textbf{9},  25118--25130 (2021). \doi{10.1109/ACCESS.2021.3056088}

\bibitem{hinton2006reducing}
Hinton, G.E., Salakhutdinov, R.R.: Reducing the dimensionality of data with neural networks. science  \textbf{313}(5786),  504--507 (2006)

\bibitem{FBCSP}
{Kai Keng Ang}, {Zheng Yang Chin}, {Haihong Zhang}, {Cuntai Guan}: Filter bank common spatial pattern ({FBCSP}) in {B}rain-{C}omputer {I}nterface. In: 2008 IEEE Int. Joint Conf. on Neural Networks (IEEE World Congress on Computational Intelligence) (2008)

\bibitem{VAE_ORIGINAL_PAPER_kingma}
Kingma, D.P., Welling, M.: Auto-encoding variational bayes. arXiv preprint arXiv:1312.6114  (2013)

\bibitem{kobler2020}
Kobler, R.J., Kolesnichenko, E., Sburlea, A.I., M{\"u}ller-Putz, G.R.: Distinct cortical networks for hand movement initiation and directional processing: an eeg study. NeuroImage  \textbf{220},  117076 (2020)

\bibitem{koelstra_2012_DEAP_dataset}
Koelstra, S., Muhl, C., Soleymani, M., Lee, J.S., Yazdani, A., Ebrahimi, T., Pun, T., Nijholt, A., Patras, I.: Deap: A database for emotion analysis ;using physiological signals. IEEE Transactions on Affective Computing  \textbf{3}(1),  18--31 (2012). \doi{10.1109/T-AFFC.2011.15}

\bibitem{EEGNet_paper}
Lawhern, V., Solon, A., Waytowich, N., Gordon, S., Hung, C., Lance, B.: {EEGN}et: A compact convolutional network for {EEG}-based {B}rain-{C}omputer {I}nterfaces. Journal of Neural Engineering  \textbf{15} (11 2016). \doi{10.1088/1741-2552/aace8c}

\bibitem{Lazurenko2018}
Lazurenko, D., Kiroy, V., Aslanyan, E., Shepelev, I., Bakhtin, O., Minyaeva, N.: Electrographic properties of movement-related potentials. Neuroscience and Behavioral Physiology  \textbf{48}(9),  1078--1087 (2018)

\bibitem{DFNN_AC_TAB_7}
Li, D., Wang, J., Xu, J., Fang, X.: Densely feature fusion based on convolutional neural networks for motor imagery {EEG} classification. IEEE Access  \textbf{7},  132720--132730 (2019). \doi{10.1109/ACCESS.2019.2941867}

\bibitem{li2018}
Li, H., Huang, G., Lin, Q., et~al.: Combining movement-related cortical potentials and event-related desynchronization to study movement preparation and execution. Frontiers in neurology  \textbf{9}, ~822 (2018)

\bibitem{li2019disentangled}
Li, Y., Pan, Q., Wang, S., Peng, H., Yang, T., Cambria, E.: Disentangled variational auto-encoder for semi-supervised learning. Information Sciences  \textbf{482},  73--85 (2019)

\bibitem{liu_2020_eeg_emotion_SAE_CNN}
Liu, J., Wu, G., Luo, Y., Qiu, S., Yang, S., Li, W., Bi, Y.: Eeg-based emotion classification using a deep neural network and sparse autoencoder. Frontiers in Systems Neuroscience  \textbf{14} (2020). \doi{10.3389/fnsys.2020.00043}, \url{https://www.frontiersin.org/articles/10.3389/fnsys.2020.00043}

\bibitem{AdamW_paper}
Loshchilov, I., Hutter, F.: Decoupled weight decay regularization. In: International Conference on Learning Representations (2019), \url{https://openreview.net/forum?id=Bkg6RiCqY7}

\bibitem{luo_2020_reconstruction_with_gan}
Luo, T.j., Fan, Y., Chen, L., Guo, G., Zhou, C.: Eeg signal reconstruction using a generative adversarial network with wasserstein distance and temporal-spatial-frequency loss. Frontiers in Neuroinformatics  \textbf{14} (2020). \doi{10.3389/fninf.2020.00015}, \url{https://www.frontiersin.org/articles/10.3389/fninf.2020.00015}

\bibitem{luo2018eeggan}
Luo, Y., Lu, B.L.: Eeg data augmentation for emotion recognition using a conditional wasserstein gan. In: 2018 40th Annual International Conference of the IEEE Engineering in Medicine and Biology Society (EMBC). pp. 2535--2538 (2018). \doi{10.1109/EMBC.2018.8512865}

\bibitem{Magnuson2021}
Magnuson, J.R., McNeil, C.J.: Low-frequency neural activity at rest is correlated with the movement-related cortical potentials elicited during both real and imagined movements. Neuroscience Letters  \textbf{742},  135530 (2021). \doi{https://doi.org/10.1016/j.neulet.2020.135530}, \url{https://www.sciencedirect.com/science/article/pii/S0304394020308004}

\bibitem{WCOM2021}
Martiradonna, S., Cisotto, G., Boggia, G., Piro, G., Vangelista, L., Tomasin, S.: Cascaded wlan-fwa networking and computing architecture for pervasive in-home healthcare. IEEE Wireless Communications  \textbf{28}(3),  92--99 (2021)

\bibitem{Monolitich_network_AC_KS_TAB_3}
Olivas, B.E., Chacon, M.: Classification of multiple motor imagery using deep convolutional neural networks and spatial filters. Applied Soft Computing  \textbf{75} (11 2018). \doi{10.1016/j.asoc.2018.11.031}

\bibitem{Pfurtscheller2006}
Pfurtscheller, G., Brunner, C., Schl{\"o}gl, A., Da~Silva, F.L.: Mu rhythm (de) synchronization and {EEG} single-trial classification of different motor imagery tasks. NeuroImage  \textbf{31}(1),  153--159 (2006)

\bibitem{yang_2018_DSAE_EEG}
Qiu, Y., Zhou, W., Yu, N., Du, P.: Denoising sparse autoencoder-based ictal eeg classification. IEEE Transactions on Neural Systems and Rehabilitation Engineering  \textbf{26}(9),  1717--1726 (2018). \doi{10.1109/TNSRE.2018.2864306}

\bibitem{MI_EEGNet_AC_TAB_10}
Riyad, M., Khalil, M., Adib, A.: {MI-EEGNET}: A novel convolutional neural network for motor imagery classification. Journal of Neuroscience Methods  \textbf{353},  109037 (2021). \doi{https://doi.org/10.1016/j.jneumeth.2020.109037}, \url{https://www.sciencedirect.com/science/article/pii/S016502702030460X}

\bibitem{Sakhavi_AC_KS_TAB_4_5}
{Sakhavi}, S., {Guan}, C., {Yan}, S.: Learning temporal information for brain-computer interface using convolutional neural networks. IEEE Transactions on Neural Networks and Learning Systems  \textbf{29}(11),  5619--5629 (2018)

\bibitem{qEEGNet}
Schneider, T., Wang, X., Hersche, M., Cavigelli, L., Benini, L.: {Q-EEGNet}: An energy-efficient 8-bit quantized parallel {EEGN}et implementation for edge motor-imagery brain-machine interfaces. In: 2020 IEEE International Conference on Smart Computing (SMARTCOMP). pp. 284--289. IEEE (2020)

\bibitem{SeelandMRCP2015}
Seeland, A., Manca, L., Kirchner, F., Kirchner, E.A.: Spatio-temporal comparison between erd/ers and mrcp-based movement prediction. In: Proceedings of the International Conference on Bio-inspired Systems and Signal Processing - Volume 1: BIOSIGNALS, (BIOSTEC 2015). pp. 219--226. INSTICC, SciTePress (2015). \doi{10.5220/0005214002190226}

\bibitem{Frontiers2014}
Silvoni, S., Cavinato, M., Volpato, C., Cisotto, G., Genna, C., Agostini, M., Turolla, A., Ramos-Murguialday, A., Piccione, F.: Kinematic and neurophysiological consequences of an assisted-force-feedback brain-machine interface training: a case study. Frontiers in neurology  \textbf{4}, ~173 (2013)

\bibitem{tian_2023_Dual_encoder_VAE_GAN_generation}
Tian, C., Ma, Y., Cammon, J., Fang, F., Zhang, Y., Meng, M.: Dual-encoder vae-gan with spatiotemporal features for emotional eeg data augmentation. IEEE Transactions on Neural Systems and Rehabilitation Engineering  \textbf{31},  2018--2027 (2023). \doi{10.1109/TNSRE.2023.3266810}

\bibitem{Wang_2020_EEGNet_autoencoder_1}
Wang, J., Wei, M., Zhang, L., Huang, G., Liang, Z., Li, L., Zhang, Z.: An autoencoder-based approach to predict subjective pain perception from high-density evoked eeg potentials. In: 2020 42nd Annual International Conference of the IEEE Engineering in Medicine and Biology Society (EMBC). pp. 1507--1511 (2020). \doi{10.1109/EMBC44109.2020.9176644}

\bibitem{zancanaro_CIBCB_article}
Zancanaro, A., Cisotto, G., Paulo, J.R., Pires, G., Nunes, U.J.: {CNN}-based approaches for cross-subject classification in motor imagery: From the state-of-the-art to {D}ynamic{N}et. In: 2021 IEEE Conference on Computational Intelligence in Bioinformatics and Computational Biology (CIBCB). pp.~1--7 (2021). \doi{10.1109/CIBCB49929.2021.9562821}

\bibitem{MetroAgriFor2022}
Zancanaro, A., Cisotto, G., Tegegn, D.D., Manzoni, S.L., Reguzzoni, I., Lotti, E., Zoppis, I.: Variational autoencoder for early stress detection in smart agriculture: A pilot study. In: 2022 IEEE Workshop on Metrology for Agriculture and Forestry (MetroAgriFor). pp. 126--130. IEEE (2022)

\bibitem{Zancanaro_2023_vEEGNet}
Zancanaro, A., Zoppis, I., Manzoni, S., Cisotto, G.: veegnet: A new deep learning model to classify and generate eeg. In: Proceedings of the 9th International Conference on Information and Communication Technologies for Ageing Well and e-Health - Volume 1: ICT4AWE,. pp. 245--252. INSTICC, SciTePress (2023). \doi{10.5220/0011990800003476}

\bibitem{zheng_2015_SEED_dataset}
Zheng, W.L., Lu, B.L.: Investigating critical frequency bands and channels for eeg-based emotion recognition with deep neural networks. IEEE Transactions on Autonomous Mental Development  \textbf{7}(3),  162--175 (2015). \doi{10.1109/TAMD.2015.2431497}

\bibitem{BIOSIGNALS2020}
Zoppis, I., Zanga, A., Manzoni, S., Cisotto, G., Morreale, A., Stella, F., Mauri, G.: An attention-based architecture for eeg classification. In: BIOSIGNALS. pp. 214--219 (2020)

\end{thebibliography}
\end{document}